\documentclass[aps,prl,twocolumn,superscriptaddress,address,amsmath,showpacs]{revtex4}
\usepackage{graphicx}

\begin{document}

%\preprint{}

\title{Effect of Surface Andreev Bound States on the Bean-Livingston Barrier
in d-Wave Superconductors}


\author{C. Iniotakis}
%\thanks{}
\affiliation{Institute for Theoretical Physics, ETH Zurich, 8093 Zurich, Switzerland}
\author{T. Dahm}
\affiliation{Institut f\"ur Theoretische Physik, Universit\"at  T\"ubingen, Auf der Morgenstelle 14, D-72076 T\"ubingen, Germany}
\affiliation{Institute for Solid State Physics, University of Tokyo, Kashiwanoha, Kashiwa,
Chiba 277-8581, Japan}
\author{N. Schopohl}
\affiliation{Institut f\"ur Theoretische Physik, Universit\"at  T\"ubingen, Auf der Morgenstelle 14, D-72076 T\"ubingen, Germany}


\date{\today}

\begin{abstract}
We study the influence of surface Andreev bound states in $d$-wave superconductors on the
Bean-Livingston surface barrier for entry of a vortex line into a strongly type-II superconductor.
Starting from Eilenberger theory we derive a generalization of London theory to
incorporate the anomalous surface currents arising from the Andreev bound states.
This allows us to find an analytical expression for the modification of the
Bean-Livingston barrier in terms of a single parameter describing the
influence of the Andreev bound states. We find that the field of first vortex
entry is significantly enhanced. Also, the depinning field for vortices
near the surface is renormalized. Both effects are temperature
dependent and depend on the orientation of the surface relative to
the $d$-wave gap.
 \end{abstract}



\maketitle

%\section{}
%\subsection{}
%\subsubsection{}

Formation and dynamics of vortices is of interest in such diverse
fields as cosmology, liquid crystals, and superfluids \cite{Kibble}.
It is well known that fermionic states existing inside
a vortex can significantly alter the structure of a vortex line
\cite{KramerPesch}. In unconventional superconductors fermionic
states, so-called Andreev bound states, can also appear at the surface.
In this work we show that such surface states 
strongly influence the entrance and pinning properties of a vortex.
Entry of a vortex into a type-II superconductor is
hindered by the so-called Bean-Livingston (BL) surface barrier \cite{Bean}.
This barrier arises due to a competition of two forces acting on
the vortex line: a force coming from the external magnetic field
driving the vortex into the superconductor and a force of the
vortex' mirror image attracting it towards the outside.
As a result, the penetration field $B_s$ of first vortex entry is typically
much larger than the lower critical field $B_{c1}$ and becomes of the
order of the thermodynamic critical field $B_c$ \cite{deGennes}.

At the surface of a $d$-wave
superconductor Andreev bound states exist, depending on
the relative orientation of the $d$-wave gap function to
the surface \cite{Hu,Tanaka95,Buchholtz,KashiwayaReport}.
It has been shown that these
states split in the presence of a supercurrent running along the
surface, generating a quasi-particle current directed opposite to
the supercurrent \cite{Fogelstroem}. At low temperatures this
anomalous Meissner current even over-compensates the supercurrent,
leading to a reversal of the current flow at the surface 
\cite{Fogelstroem,Walter}. Even though the Andreev bound states only
exist within a coherence length $\xi$ from the surface, they are influenced
by a vortex line already, when the vortex is a penetration depth $\lambda \gg \xi$
away from the surface, because the supercurrent field around the
vortex is long-ranged \cite{Vortexshadow}. Therefore, a modification
of the BL barrier has to be expected due to the change
of the free energy of the Andreev bound states as a function of the
position of the vortex creating an additional force. Here, we study 
this modification of the BL barrier and show that it
results in a sizeable increase of the penetration field and
the depinning field.

We consider a superconducting half-space in the region $x\geq 0$.
Both the cylindrical Fermi surface of the $d$-wave superconductor and the external magnetic
field $\mathbf{B}_0$ shall be aligned parallel to the $z$-axis. In the following the single 
Abrikosov vortex 
is situated at the position $\mathbf{r}_v=(x_v,0)$ on the $x$-axis, and the specular surface
shall be given by the $y$-axis.
Since we wish to explore Andreev bound states in an inhomogenous
superconductor, Ginzburg-Landau theory is not sufficient and as a 
starting point the minimum theory necessary is Eilenberger theory
\cite{Eilenberger,Larkin,Serene}.
In particular, we use the Riccati-parametrization \cite{Schopohl,Eschrig},
where local equilibrium properties of the $d$-wave superconductor are determined 
in terms of two scalar coherence functions $a$ and $b$. To each  
unit vector $\hat{\mathbf{k}}$ parametrizing
the Fermi surface of the superconductor corresponds a 
quasiparticle trajectory in real space which is parallel to 
the Fermi velocity $\mathbf{v}_F (\hat{\mathbf{k}})$.
Along this trajectory, and for given Matsubara  
frequency $\varepsilon_n=(2n+1)\pi k_B T$, 
two decoupled differential equations of the Riccati type have to be integrated
\begin{eqnarray}
\label{EQRiccati1}
\langle \hbar \mathbf{v}_F, \nabla \rangle a + (2\varepsilon_n -2ie/c  \langle \mathbf{v}_F,\mathbf{A}\rangle+\Delta^* a)a -\Delta &=&0 \\
\nonumber
\langle \hbar \mathbf{v}_F, \nabla \rangle b - (2\varepsilon_n -2ie/c \langle \mathbf{v}_F,\mathbf{A}\rangle+\Delta b)b +\Delta^* &=&0. 
\end{eqnarray}
Here, the arguments have been omitted for brevity, and in the
following we use the specific gauge, in which
the gap function $\Delta$ is real and a phase gradient is absorbed
into the magnetic vector potential $\mathbf{A}$.
Starting from known bulk values,  solutions for $a$ ($b$) can be calculated  in a stable way numerically, if
the integration of the corresponding Riccati equation is performed parallel (antiparallel) to the trajectory direction.
Both the gap function $\Delta(\mathbf{r},\hat{\mathbf{k}})=\Delta_\infty \chi(\hat{\mathbf{k}}) \psi(\mathbf{r})$
 and the magnetic vector potential $\mathbf{A}(\mathbf r)$ 
can be regarded as spatially dependent input parameters for the 
Riccati equations (\ref{EQRiccati1}). 
In general, however, their correct selfconsistent form  has to be determined from two additional conditions.
One of them is given by the gap equation
\begin{equation}
\label{EQGap}
\Delta_\infty \psi(\mathbf r) = 2\pi N_0 V k_B T  \sum^{\omega_c}_{\varepsilon_n >0} \left\langle \chi(\hat{\mathbf k}) \frac{2a}{1+ab} (\mathbf r,\hat{\mathbf k},\varepsilon_n)\right\rangle,
\end{equation} 
where $N_0$ is the density of states in the normal phase, $V$  the interaction strength,  
$\omega_c$  the cut-off energy and $\langle...\rangle$ denotes an average over the Fermi 
surface of the superconductor. Furthermore, the momentum 
dependence of the $d$-wave gap is implemented by
$\chi(\hat{\mathbf k})=\cos 2\phi$. 
The other condition regards electrodynamic selfconsistency. It can be ensured 
via the quasiclassical expression for the current density
\begin{equation}
\label{EQjQC}
\mathbf{j}(\mathbf r)=-4\pi i e N_0 k_B T \sum^{\omega_c}_{\varepsilon_n > 0} \left\langle \mathbf{v}_F(\hat{\mathbf k}) \frac{1-a b}{1+a b}(\mathbf r,\hat{\mathbf k},\varepsilon_n) \right\rangle
\end{equation}
together with the Maxwell equation
\begin{equation}
\label{EQrotrot}
\nabla\times\nabla\times \mathbf{A}(\mathbf r)=\frac{4\pi}{c} \mathbf{j}(\mathbf r).
\end{equation}
The system of equations (\ref{EQRiccati1})-(\ref{EQrotrot}) is  self-contained. 
In principle it can be used to find a fully selfconsistent solution, which is very time-consuming,
however.

\begin{figure}[t]
\includegraphics[width=0.82 \columnwidth]{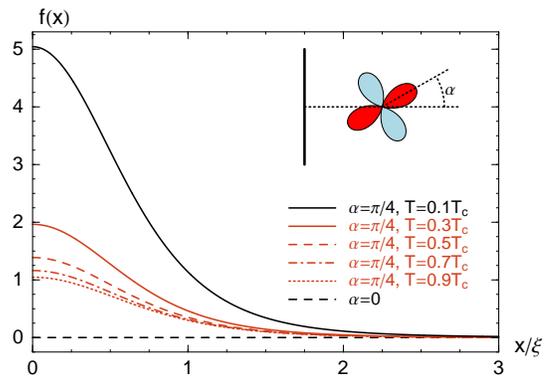}
\caption{\label{Fig01} (Color online)
Dimensionless function $f$, determining the local 
current response according to  Eq. (\ref{EQjrho}), as a function
of the distance $x$ to the boundary of a $d$-wave superconductor. 
Curves are shown for different temperatures and orientations.
London electrodynamics, i.e. $f=0$, holds  for the $d$-wave orientation $\alpha=0$ and
towards the bulk.  
For  $\alpha=\pi /4$  the current response  in the vicinity of  the 
boundary is modified due to surface Andreev bound states. This effect
is strongly enhanced for lower temperatures.
}
\end{figure}

In the following we present a way to derive a very good approximation to
the fully selfconsistent solution of Eqs. (\ref{EQRiccati1})-(\ref{EQrotrot}).
We concentrate on a $d$-wave superconductor with $\kappa=\lambda / \xi \gg 1$.
Furthermore the superconductor is considered to be clean, but not superclean. 
In other words, the effective mean free path $v_F \tau$ along a quasiparticle trajectory 
may be larger than the coherence length $\xi=\hbar v_F/\Delta_\infty$, but it should also hold
$\lambda \gg v_F \tau$.
As a starting point we then use the selfconsistent solution of Eqs. (\ref{EQRiccati1})-(\ref{EQGap})
in the absence of any magnetic vector potentials and currents. This solution can be found numerically
comparatively easy since the problem is translationally invariant along the boundary.
This solution already contains both the existence of low-energy Andreev bound states and 
the local suppression of the gap in the vicinity of the boundary.
In a next step, we start perturbing this known selfconsistent solution by 
a  magnetic vector potential. Due to the assumptions stated above, 
we can restrict ourselves to perturbations, which typically vary on the 
length scale $\lambda$ and are effectively homogeneous along the mean free path
$v_F \tau$. In that case, the Riccati equations (\ref{EQRiccati1})
can be evaluated for a constant  Doppler shift in energy, $\frac{e}{c} \langle \mathbf{v}_F, \mathbf {A} \rangle$,
along the quasiparticle trajectory. Let $\alpha$ denote the angle between the surface normal
and a maximum gap direction of the $d$-wave (cf. inset of Fig. \ref{Fig01}).
For the two most symmetric cases $\alpha=0$ and $\alpha=\pi /4$ respectively, 
the expansion of Eq. (\ref{EQjQC}) to first order of the perturbing vector potential
$\mathbf {A}$ yields an equation of the form
%a thorough 
%investigation of the selfconsistency equations (\ref{EQGap}) and (\ref{EQjQC})  shows that, in first order of the 
%perturbing vector potential $\mathbf {A}$, the
%gap remains unchanged, whereas the current density is given by 
\begin{equation}
\label{EQjrho}
\mathbf{j}(\mathbf{r})=-\frac{c}{4\pi} \frac{1}{\lambda^2} [1-f(\mathbf{r})] \mathbf{A}(\mathbf{r}),
\end{equation}   
where the dimensionless function $f$ on the righthand side is independent of $\lambda$ and can 
be determined
numerically from the known fully selfconsistent solution for  $\mathbf{A}=0$ mentioned above.
%The relation between current and magnetic vector potential, generally  
%of a nonlocal nature, has become local now. 

Eq. (\ref{EQjrho}) presents a generalization of London theory including
surface effects. Physically, the influence of both Andreev bound states and a
gap  suppression  at the surface on the electrodynamics of the superconductor are
completely represented by the function $f$.
Due to translational symmetry $f$ only depends on the distance $x$ to the
boundary. 
In Fig. \ref{Fig01} we show our results for $f(x)$. 
For the $d$-wave orientation $\alpha=0$ neither Andreev bound states 
nor a suppression of the gap exist at the boundary and
$f$ vanishes identically.
In this case London electrodynamics is valid everywhere
in the superconducting half-space. For the orientation 
$\alpha=\pi /4$ however, there are significant changes of the current
response close to the boundary, since the transport is dominated  
by anomalous Meissner currents of the zero-energy Andreev bound states. The net current response $1-f$ 
may even become negative, and this effect is enhanced for 
lower temperatures since the spectral weight of  Andreev bound states 
strongly increases \cite{Fogelstroem,Walter}. A few coherence lengths
into the bulk London electrodynamics is recovered.
An important aspect of the resulting Eq. (\ref{EQjrho}) is,
that the thermodynamic and electrodynamic selfconsistency
conditions
are  effectively decoupled. As mentioned above, the function $f$ can be obtained
using the gap equation (\ref{EQGap}) only, without any electrodynamics considered. 
On the other hand, once  $f$  is provided, the magnetic vector potential 
$\mathbf{A}$ of the fully selfconsistent 
problem may simply be found from Eq. (\ref{EQrotrot}) and
the modified London equation (\ref{EQjrho}), together with Neumann
boundary conditions. 

In the following, we show how to find the 
resulting  magnetic field and derive the modified surface barrier. 
For this purpose it is useful to employ the subscript notation 
$\mathbf{B}=\mathbf{B}_L+\mathbf{B}_R$,
where the subscript $L$ refers to the well-known London solution of
the problem, whereas $R$ denotes the unknown small 'response' term
generated by the function $f$ in Eq.~(\ref{EQjrho}) due to the presence of Andreev bound states. 
The London solution itself is given by $\mathbf{B}_{L} =\mathbf{B}_{L}^W +\mathbf{B}_{L}^V $
with
%\begin{eqnarray}
%\mathbf{B}_{L}^{W} (\mathbf{r}) &= &B_0 e^{-x/\lambda} \mathbf{z} \\
%\mathbf{B}_{L}^{V} (\mathbf{r}) &=&\frac{\Phi_0}{2\pi \lambda^2} \left[ K_0 \left(\frac{ |\mathbf{r}-\mathbf{r}_v|}{\lambda} \right)  
%-K_0 \left(\frac{ |\mathbf{r}-\mathbf{r}_{v}^{*}|}{\lambda} \right) \right]  \mathbf{z},
%\nonumber
%\end{eqnarray}
$\mathbf{B}_{L}^{W} (\mathbf{r}) = B_0 e^{-x/\lambda} \mathbf{z}$ and
$\mathbf{B}_{L}^{V} (\mathbf{r}) =\frac{\Phi_0}{2\pi \lambda^2} 
\left[ K_0 \left(\frac{ |\mathbf{r}-\mathbf{r}_v|}{\lambda} \right)  
-K_0 \left(\frac{ |\mathbf{r}-\mathbf{r}_{v}^{*}|}{\lambda} \right) \right]  \mathbf{z}$
where $\Phi_0$ is the flux quantum, $K_0$ is a modified Bessel function and the star operator 
shall define a mirror operation with respect to the boundary, i.e. $\mathbf{r}^*=(-x,y)$. 
Via the superscript notation a field is uniquely divided
 into two parts. The superscript $W$ refers to the 'wall' part, which is translationally invariant along the boundary
and is due to the screened external magnetic field. The 'vortex' part denoted by  $V$ incorporates the fields
due to vortex and its mirror antivortex.
The corresponding magnetic vector potential 
of the London solution is found from $\mathbf{A}_{L}=-\lambda^2 \nabla \times \mathbf{B}_{L}$. 
Then, Eqs. (\ref{EQrotrot}) and (\ref{EQjrho}) may be combined
to 
\begin{equation}
\label{EQBrLaplace}
\left( \Delta-\frac{1}{\lambda^2} \right) \mathbf{B}_R (\mathbf{r}) =-\frac{1}{ \lambda^2} \nabla \times 
\left[ f(x) \mathbf{A}(\mathbf{r}) \right].
\end{equation} 
Inversion of the differential operator leads to an integration over the superconducting area $S$
that can be written in the form
%\begin{eqnarray}
%\label{EQBr1}
%\mathbf{B}_R (\mathbf{r}) &=&- \frac{1}{\lambda^2} \int_S d^2 r' G_S (\mathbf{r},\mathbf{r}') 
%\nabla' \times \left[ f(x') \mathbf{A}(\mathbf{r}') \right] \\
%&=&\frac{1}{\lambda^2} \int_S d^2 r'  f(x') \left[\nabla' G_S (\mathbf{r},\mathbf{r}') \right] 
% \times \mathbf{A}(\mathbf{r}') \nonumber,
%\end{eqnarray}
\begin{equation}
\label{EQBr1}
\mathbf{B}_R (\mathbf{r}) 
=\frac{1}{\lambda^2} \int_S d^2 r'  f(x') \left[\nabla' G_S (\mathbf{r},\mathbf{r}') \right] 
 \times \mathbf{A}(\mathbf{r}') ,
\end{equation}
where  the appropriate Green's function $G_S$ of the 
problem has been involved, which is given by
$
G_S (\mathbf{r},\mathbf{r}')=-\frac{1}{2\pi}\left[
K_0 \left(\frac{ |\mathbf{r}-\mathbf{r}'|}{\lambda} \right)  
-K_0 \left(\frac{ |\mathbf{r}-\mathbf{r}'^{*}|}{\lambda} \right)
 \right].
$
In particular, the unknown magnetic field ${\mathbf B}_R$ according to Eq. (\ref{EQBr1}) fulfils Eq. (\ref{EQBrLaplace}), 
is divergence-free and vanishes everywhere at the boundary. The integrand in Eq. (\ref{EQBr1}) is weighted by the 
function $f$, which is effectively non-zero only up to a finite
distance $l_f \ll \lambda$ from the boundary, where the standard London theory
is not valid (cf. Fig. \ref{Fig01}). Furthermore, also the total deviation from
standard London theory is small, i.e. $\int dx \, f \ll \lambda$. Thus, concentrating on the main
contribution to ${\mathbf B}_R$, the source term $\mathbf A$ in the integrand can 
be replaced by the London solution $\mathbf{A}_L$. Moreover it is sufficient to keep the dominating term only:
$
%\begin{equation}
%\label{EQBR}
B_R (\mathbf{r})=\frac{1}{\lambda^2} \int_S d^2 r' f(x') \left[ \partial_{x'} G_S (\mathbf{r},\mathbf{r}')  \right] A_{L,y} (\mathbf{r}').
%\end{equation}
$
This formula allows to calculate, how  the magnetic field deviates 
from the London solution due to the presence of  Andreev bound states.

Let us derive the modified BL barrier now.
The corresponding free energy density $g$ of our system is
given by
%\begin{eqnarray}
%\nonumber
%g&=&g_0+ \frac{1}{8\pi}\left[ \frac{1}{\lambda^2} (1- f) \mathbf{A}^2+\mathbf{B}^2 -2\mathbf{B}_0 \cdot \mathbf{B}  \right] \\
%&=&g_0+\frac{1}{8\pi} \nabla \cdot \left[ \mathbf{A} \times (\mathbf{B}-2\mathbf{B}_0) \right].
%\end{eqnarray}
%\begin{equation}
$
g=g_0+ \frac{1}{8\pi}\left[ \frac{1}{\lambda^2} (1- f) \mathbf{A}^2+\mathbf{B}^2 -2\mathbf{B}_0 \cdot \mathbf{B}  \right].
$
%\end{equation}
Here, $g_0$ denotes the value in the absence of magnetic fields and currents.
A rather long but straightforward analysis yields the resulting total free energy $G$ (per unit length 
of the $z$-direction) as a function of the vortex distance $x_v$:
\begin{eqnarray}
\label{EQFPre}
\nonumber
G(x_v)&=&\frac{\Phi_0}{8 \pi} \left[ 2 B_0 e^{-x_v /\lambda} -\frac{\Phi_0}{2\pi \lambda^2 } K_0(2x_v / \lambda) \right] \\
&+&\frac{\Phi_0}{8 \pi} \left[ B_R (\mathbf{r}_v) - \frac{B_0}{\Phi_0} \int_S d^2 r B^V_R (\mathbf{r}) \right]. 
\end{eqnarray}  
Here, an arbitrary additive constant has been chosen such that $G(x_v \rightarrow \infty) =0$. Note, that the first two terms
are the original BL barrier (cf. \cite{Bean}). The latter terms are modifications which only appear when
$f \neq 0$ and thus  $B_R \neq 0$, correspondingly. 
% With help of Eq. (\ref{EQBR}) these additional terms can readily be evaluated. 
After some algebra, we find
\begin{eqnarray*}
B^W_R(\mathbf{r}_v)&=& \frac{B_0}{2\lambda} e^{-x_v / \lambda} \int_0^\infty dx' f(x') \left( 1+e^{-2x' /\lambda} \right) \\
B^V_R(\mathbf{r}_v) &=&-\frac{\Phi_0}{4\pi \lambda^3} \int_0^\infty dx' f(x') \tilde{K}(x_v,x'),
\end{eqnarray*}
and $B^W_R(\mathbf{r}_v)=- \frac{B_0}{\Phi_0} \int_S d^2 r B^V_R (\mathbf{r})$
where the abbreviation
$\tilde{K}(x_v,x') =K_1 \left( 2\frac{x_v-x'}{\lambda}\right) +
2 K_1 \left( \frac{2x_v}{\lambda} \right) +
K_1 \left( 2\frac{x_v+x'}{\lambda}\right) 
$
has been used and we restricted ourselves to $x_v > l_f$, i.e. the vortex is not directly situated in the small 
surface region with $f \neq 0$.  Since the function $f$ is nonzero only in a small surface region,
the above expressions can be further simplified, if one replaces $f(x')$ by a
delta function $ c_f \delta(x')$ with 
$
c_f =\int^\infty_0 dx' f(x').
$
The quantity $c_f$ has the dimension of a length and remains the only
parameter describing the influence of the Andreev bound states.
Then, the final result for the normalized free energy of Eq. (\ref{EQFPre}) is
\begin{equation}
\label{EQF}
\hat{G}(x_v) =   2 \hat{B}_0 \left( 1+\hat{c}_f \right) e^{-\hat{x}_v} 
- \left[ K_0(2\hat{x}_v) +2 \hat{c}_f K_1(2\hat{x}_v) \right] .
\end{equation}
Here, the normalized $\hat{G}$ is defined as  $G/(\Phi_0^2 /16\pi^2 \lambda^2)$,  
$\hat{B}_0$ denotes the applied magnetic field in units of  
$ \Phi_0/2\pi \lambda^2$, and we abbreviate $\hat{x}_v =x_v /\lambda$
and  $\hat{c}_f =c_f /\lambda$. 

Eq. (\ref{EQF}) is our generalization of the surface barrier. For $\hat{c}_f=0$
it reduces to the result of Bean and Livingston \cite{Bean,deGennes}. In the
presence of Andreev bound states, however, there are two modifications
due to $\hat{c}_f \neq0$. The first modification is an enhancement of the
repulsive part of the potential, effectively increasing the external magnetic 
field amplitude. Secondly, also the attractive part of the potential
coming from the antivortex is strengthened by an additional term.
Note, that $\hat{c}_f \ll1$. Even the maximum value of Fig. \ref{Fig01}
occuring for the orientation $\alpha=\pi/4$ at the temperature $T=0.1 T_c$
is only $c_f \approx 3.71 \xi$, i.e.  $\hat{c}_f \approx 3.71 \kappa^{-1}$.
Nevertheless particularly the second term becomes important, because
it contains a Bessel function, which diverges for small vortex distances  
$x_v$. 

\begin{figure}[t]
\includegraphics[width=0.82 \columnwidth]{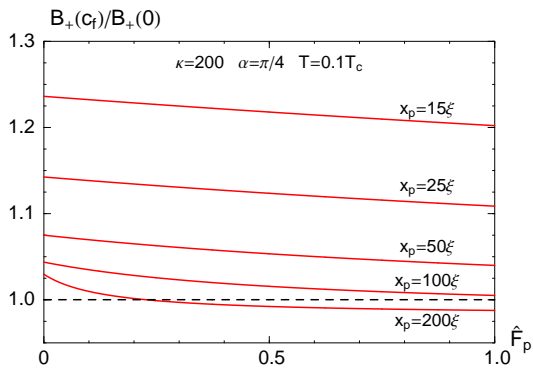}
\caption{\label{Fig02} (Color online)
Relative change of the external magnetic field $B_+$ needed to release  
a vortex from a pinning center  towards the bulk. Results are shown 
as a function of the normalized pinning force
$\hat{F}_p$ for different positions $x_p$ of the pinning center. 
Obviously, unpinning the vortex  in the presence of surface bound states 
requires a significantly higher external
magnetic field.
}
\end{figure}

It is important to realize, how drastic some properties may change
just because of this rather small looking modification. To give an example
we consider a vortex  pinned in a pinning center at the
distance $x_p$  from the boundary.
The maximum pinning force of that center, measured in 
units of $\Phi_0^2/8\pi^2 \lambda^3$, is characterized by the parameter $\hat{F}_p$. 
If the applied magnetic field then exceeds some specific value  $B_+$, which can be 
determined from Eq. (\ref{EQF}), the vortex gets unpinned and proceeds towards
the bulk. The relative change for this depinning field is given by
\begin{equation}
\frac{B_+ (c_f)}{B_+ (0)}=\frac{1+\frac{K_1 (2\hat{x}_p)}{K_1 (2\hat{x}_p) + \hat{F}_p} 
\left( 2 \hat{c}_f \frac{K_0 (2\hat{x}_p)}{K_1 (2\hat{x}_p)} +\frac{c_f}{x_p}  \right)}{1+\hat{c}_f}.
\end{equation}
For the parameters $\kappa=200$ and $c_f=3.71 \xi$ 
from above, some results are shown in Fig. \ref{Fig02}. Clearly, the external magnetic field needed to unpin a vortex 
close to a boundary which exhibits Andreev bound states
is significantly enhanced in the most relevant cases.

\begin{figure}[t]
\includegraphics[width=0.82 \columnwidth]{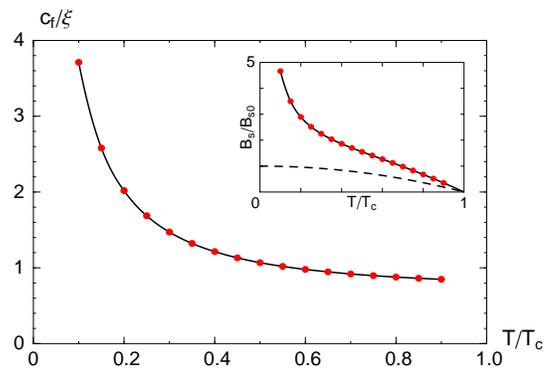}
\caption{\label{Fig03} (Color online)
Temperature dependence of the parameter $c_f/\xi$ for $\alpha=\pi/4$.
The inset shows the expected temperature dependence of the
field of first vortex entry $B_s$ according to Eq.~(\ref{entryfieldapprox})
(solid line) compared with the BL result (dashed line).
}
\end{figure}

Analogous to the classical BL work \cite{Bean}, we can roughly 
estimate the field of first free vortex entry $B_s$
by choosing $\hat{F}_p= 0$ and
a distance $x_v \sim \xi$ from the boundary. Then, we find
\begin{equation}
\frac{B_s(c_f)}{B_s (0)} \approx 1 +\frac{c_f}{\xi}.
\label{entryfieldapprox}
\end{equation}
where we have assumed $\hat{c}_f \ll 1$, but $c_f/\xi \sim 1$.
Here, $B_s (0)$ is the BL value.
This result means that the field of first vortex entry is significantly
enhanced by the presence of Andreev bound states.
Compared to the classical BL value it may increase
by a factor 4-5 for low temperatures and $\alpha=\pi/4$. In Fig.~\ref{Fig03} 
we show the temperature dependence of $c_f/\xi$. 
The corresponding characteristic change of  $B_s$ as
a function of temperature, which is indicated by the inset of Fig.~\ref{Fig03}, may 
facilitate experimental observation of this effect.

We also expect the modification of the BL barrier
to have a strong influence on the hysteresis of the split of 
the zero bias conductance peak as observed in
tunneling experiments \cite{Aprili,Krupke}. However,
a detailed calculation of the hysteresis curve goes beyond 
the scope of the present work.

\end{document}